\documentstyle[12pt]{article}

\catcode`\@=11
\long\def\@makefntext#1{
\protect\noindent \hbox to 3.2pt {\hskip-.9pt  
$^{{\ninerm\@thefnmark}}$\hfil}#1\hfill}		

\def\@makefnmark{\hbox to 0pt{$^{\@thefnmark}$\hss}}  
	
\def\ps@myheadings{\let\@mkboth\@gobbletwo
\def\@oddhead{\hbox{}
\rightmark\hfil\ninerm\thepage}   
\def\@oddfoot{}\def\@evenhead{\ninerm\thepage\hfil
\leftmark\hbox{}}\def\@evenfoot{}
\def\sectionmark##1{}\def\subsectionmark##1{}}

\renewcommand{\thefootnote}{\fnsymbol{footnote}}

\newcounter{sectionc}\newcounter{subsectionc}\newcounter{subsubsectionc}
\renewcommand{\section}[1] {\vspace*{0.6cm}\addtocounter{sectionc}{1} 
\setcounter{subsectionc}{0}\setcounter{subsubsectionc}{0}\noindent 
	{\normalsize\bf\thesectionc. #1}\par\vspace*{0.4cm}}
\renewcommand{\subsection}[1] {\vspace*{0.6cm}\addtocounter{subsectionc}{1} 
	\setcounter{subsubsectionc}{0}\noindent 
	{\normalsize\it\thesectionc.\thesubsectionc. #1}\par\vspace*{0.4cm}}
\renewcommand{\subsubsection}[1]
{\vspace*{0.6cm}\addtocounter{subsubsectionc}{1}
	\noindent {\normalsize\rm\thesectionc.\thesubsectionc.\thesubsubsectionc. 
	#1}\par\vspace*{0.4cm}}

\newcounter{appendixc}
\newcounter{subappendixc}[appendixc]
\newcounter{subsubappendixc}[subappendixc]

\renewcommand{\appendix}[1] {\vspace*{0.6cm}
        \refstepcounter{appendixc}
        \setcounter{figure}{0}
        \setcounter{table}{0}
        \setcounter{equation}{0}
        \renewcommand{\thefigure}{\Alph{appendixc}.\arabic{figure}}
        \renewcommand{\thetable}{\Alph{appendixc}.\arabic{table}}
        \renewcommand{\theappendixc}{\Alph{appendixc}}
        \renewcommand{\theequation}{\Alph{appendixc}.\arabic{equation}}
        \noindent{\bf Appendix \theappendixc #1}\par\vspace*{0.4cm}}

\def\abstracts#1{{
	\centering{\begin{minipage}{12.2truecm}\footnotesize\baselineskip=12pt\noindent
	\centerline{\footnotesize ABSTRACT}\vspace*{0.3cm}
	\parindent=0pt #1
	\end{minipage}}\par}} 

\renewenvironment{thebibliography}[1]
	{\begin{list}{\arabic{enumi}.}
	{\usecounter{enumi}\setlength{\parsep}{0pt}
\setlength{\leftmargin 1.25cm}{\rightmargin 0pt}
	 \setlength{\itemsep}{0pt} \settowidth
	{\labelwidth}{#1.}\sloppy}}{\end{list}}

\newcounter{itemlistc}
\newcounter{romanlistc}
\newcounter{alphlistc}
\newcounter{arabiclistc}

\newcommand{\fcaption}[1]{
        \refstepcounter{figure}
        \setbox\@tempboxa = \hbox{\footnotesize Fig.~\thefigure. #1}
        \ifdim \wd\@tempboxa > 6in
           {\begin{center}
        \parbox{6in}{\footnotesize\baselineskip=12pt Fig.~\thefigure. #1}
            \end{center}}
        \else
             {\begin{center}
             {\footnotesize Fig.~\thefigure. #1}
              \end{center}}
        \fi}

\newcommand{\tcaption}[1]{
        \refstepcounter{table}
        \setbox\@tempboxa = \hbox{\footnotesize Table~\thetable. #1}
        \ifdim \wd\@tempboxa > 6in
           {\begin{center}
        \parbox{6in}{\footnotesize\baselineskip=12pt Table~\thetable. #1}
            \end{center}}
        \else
             {\begin{center}
             {\footnotesize Table~\thetable. #1}
              \end{center}}
        \fi}

\def\@citex[#1]#2{\if@filesw\immediate\write\@auxout
	{\string\citation{#2}}\fi
\def\@citea{}\@cite{\@for\@citeb:=#2\do
	{\@citea\def\@citea{,}\@ifundefined
	{b@\@citeb}{{\bf ?}\@warning
	{Citation `\@citeb' on page \thepage \space undefined}}
	{\csname b@\@citeb\endcsname}}}{#1}}

\newif\if@cghi
\def\cite{\@cghitrue\@ifnextchar [{\@tempswatrue
	\@citex}{\@tempswafalse\@citex[]}}
\def\citelow{\@cghifalse\@ifnextchar [{\@tempswatrue
	\@citex}{\@tempswafalse\@citex[]}}
\def\@cite#1#2{{$\null^{#1}$\if@tempswa\typeout
	{IJCGA warning: optional citation argument 
	ignored: `#2'} \fi}}

 1
 1
 1

\font\ninerm=cmr9

\begin{document}
\ifx\href\undefined\else\endinput\fi

\noindent
\hspace*{6.5cm}
YUMS 97-018, DO-TH 97-15, SNUTP 97-089 \\
\noindent
\hspace*{6.5cm}
hep-ph/9706451 (modified 27 June 1997)\\

\vspace{0.5cm}

\centerline{\normalsize\bf FLAVOR DEMOCRACY AND QUARK MASS MATRICES
\footnote{Talk given by C.S. Kim at the Workshop on Masses and Mixings 
of Quarks and Leptons, Shizuoka, Japan, March 19-21, 1997. 
Proceedings will be published.}}
\vspace*{0.3cm}

\centerline{\footnotesize C.~S.~ Kim}
\baselineskip=13pt
\centerline{\footnotesize\it Department of Physics, Yonsei University, 
Seoul 120-749, Korea}
\baselineskip=12pt
\centerline{\footnotesize E-mail: kim@cskim.yonsei.ac.kr}
\vspace*{0.3cm}
\centerline{\footnotesize and}
\vspace*{0.3cm}
\centerline{\footnotesize G.~ Cveti\v c}
\baselineskip=13pt
\centerline{\footnotesize\it Department of Physics, 
University of Dortmund, Dortmund, Germany}
\baselineskip=12pt
\centerline{\footnotesize E-mail: cvetic@doom.physik.uni-dortmund.de}

\vspace*{0.9cm}
\abstracts{Based on experimental mass hierarchy,
a set of flavor--democratic (FD) quark mass matrices at low energies 
is discussed. The model predicts
CP violation parameters $J_{CP} = (0.3 \pm 0.2)~10^{-4}$
and
${\varepsilon'/\varepsilon}  = (0.6 \pm 0.5)~10^{-3}$.
However, this simple FD model
also predicts a physical top quark mass not much higher than $100$ GeV.
As a next step, we assume that the Standard Model (SM) breaks down around 
some high energy $\Lambda$, and is replaced by a new FD flavor 
gauge theory (FGT). 
This possibility can be investigated by studying renormalization group
equations for the Yukawa couplings of  SM  with two  Higgs doublets for
various  $m_t$  and $v_{_U}/v_{_D}$.  With  appropriate flavor--democratic
boundary  conditions at $\Lambda_{\rm FGT}$, bounds on masses
of top quark and tau-neutrino are derived, 
which are compatible with experimental bounds.}
 
\normalsize\baselineskip=15pt
\setcounter{footnote}{0}
\renewcommand{\thefootnote}{\alph{footnote}}
\section{Flavor Democracy at Low Energy}
In the standard electroweak theory, the hierarchical pattern of the quark 
masses and their mixing remains an outstanding issue.
While a gauge interaction is characterized by its universal coupling constant,
the Yukawa interactions have as many coupling constants as there are fields
coupled to the Higgs boson.
There is no apparent underlying principle which governs the hierarchy of the
various Yukawa couplings,
and as a result,
the Standard Model of strong and electroweak interactions
can predict neither the quark (or lepton) masses nor their mixing.
This situation can be improved 
by assuming a universal Yukawa interaction --
the resulting spectrum consists then of one massive 
and two massless quarks
in each (up and down) sector in the three generation Standard Model.
Flavor--democratic (FD) quark mass matrices, and a perturbed form
of such FD matrices, were introduced already in 1978 by Harari,
Haut and Weyers\cite{0)} in a left-right symmetric framework.
Flavor democracy has recently been suggested by Koide, 
Fritzsch and Plankl\cite{1)},
as well as Nambu\cite{[3]} and many other authors\cite{[3]} 
as an analogy with the BCS theory of superconductivity.
In this Section we will discuss how this flavor symmetry
can be broken by a slight perturbation at low energies,
in order to reproduce the quark masses and the CKM matrix\cite{3)}.
As a result, predictions for the top quark mass
and for the CP violation parameter $J_{CP}$ are obtained.
This Section is based on a work by Cuypers and Kim\cite{11)}.

Considering only quark fields, the gauge invariant Yukawa Lagrangian  is
\begin{equation}
{\cal L}_{\rm Y} = 
- \sum_{i,j} (\bar Q'_{iL}~\Gamma^D_{ij}~d'_{jR}~\phi~+~
\bar Q'_{iL}~\Gamma^U_{ij}~u'_{jR}~\tilde \phi~+~\mbox{h.c.}) \ .
\label{eq1}
\end{equation}
Here, the primed quark fields are in a flavor [$SU(2)$] basis
of the $SU(2) \times U(1)$ electroweak gauge group --
the left-handed quarks form doublets under the $SU(2)$ transformation,
$\bar Q'_L=(\bar u'_L,~\bar d'_L)$, and the right-handed quarks are singlets.
The indices $i$ and $j$ run over the number of fermion generations.
The Yukawa coupling matrices $\Gamma^{U,D}$ are arbitrary
and not necessarily diagonal.
After spontaneous symmetry breaking, the Higgs field $\phi$  acquires 
a nonvanishing vacuum expectation value (VEV) $v$
which yields quark mass terms in the original Lagrangian
\begin{equation}
{\cal L}_{\rm mass} = - \sum_{i,j} (\bar d'_{iL}~M^D_{ij}~
d'_{jR}~+~\bar u'_{iL}~M^U_{ij}~u'_{jR}~+~\mbox{h.c.}) 
\ ,
\label{eq2}
\end{equation}
and the quark mass matrices are defined as
\begin{equation}
M^{U,D}_{ij} \equiv {v \over \sqrt{2}}~\Gamma^{U,D}_{ij}
\ .
\label{eq3}
\end{equation}
Mass matrices $M^{U,D}$ are diagonalized by biunitary
transformations involving
unitary matrices $U^{U,D}_L$ and $U^{U,D}_R$,
and the flavor eigenstates are tranformed to physical mass eigenstates
by the same unitary transformations,
\begin{equation}
U^{U,D}_L~M^{U,D}~(U^{U,D}_R)^{\dagger} = M^{U,D}_{\rm diag}~~{\rm and}~~
U^U_{L,R}~u'_{L,R} = u_{L,R},~~U^D_{L,R}~d'_{L,R} = d_{L,R}~~.
\label{eq4}
\end{equation}
Using the recent CDF data\cite{4)} of the physical top mass 
$m_t^{\rm phys.} \approx 175$ GeV,
the diagonalized mass matrices $M^{U,D}_{\rm diag}$ 
at a mass scale of 1 GeV are
\begin{equation}
M_{\rm diag}^U \approx m_t
\left[ \begin{array}{ccc}
2.5\times10^{-5} & & \\
 & 0.006 & \\
 & & 1 
\end{array} \right]
\quad {\rm and} \quad
M_{\rm diag}^D \approx m_b
\left[ \begin{array}{ccc}
 1.7\times10^{-3} & & \\
 & 0.03 & \\
 & & 1 
\end{array} \right].
\label{eq5}
\end{equation}
The first two eigenvalues in both matrices are almost zero
(almost degenerate) when compared to the 
eigenvalue of the third generation. 
In order to account for this large mass gap,
one can use mass matrices which have in a flavor basis
the flavor--democratic (FD) form
\begin{equation}
M^U_0 = \frac{m_t}{3}
\left[ \begin{array}{ccc}
1 & 1 & 1 \\
1 & 1 & 1 \\
1 & 1 & 1
\end{array} \right]
~~{\rm and}~~
M^D_0 = \frac{m_b}{3}
\left[ \begin{array}{ccc}
1 & 1 & 1 \\
1 & 1 & 1 \\
1 & 1 & 1
\end{array} \right]
~~.
\label{eq6}
\end{equation}
Diagonalization leads to a pattern
similar to the experimental spectrum (5)
\begin{equation}
M_{\rm diag}^U = m_t
\left[ \begin{array}{ccc}
0 &   &  \\
  & 0 &  \\
  &   & 1
\end{array} \right]
\qquad {\rm and} \qquad
M_{\rm diag}^D = m_b
\left[ \begin{array}{ccc}
0 &   &   \\
  & 0 &   \\
  &   & 1
\end{array} \right]
\ .
\label{eq7}
\end{equation}
Arbitrariness in the choice of the Yukawa Lagrangian
has been substantially reduced with this symmetric choice.
Each (up or down) quark sector is determined
in this pure FD approximation
by a single universal Yukawa coupling.

To induce nonzero masses for the lighter quarks
and to reproduce the experimental CKM matrix,
small perturbations have to be added to the universal Yukawa interactions.
One possibility is to analyze effects of the following
two kinds of independent perturbation matrices
\begin{equation}
P_1 =
\left[ \begin{array}{ccc}
\alpha & 0 & 0 \\
0 & \beta & 0 \\
0 & 0 & 0 
\end{array} \right]
\qquad {\rm and} \qquad
P_2 =
\left[ \begin{array}{ccc}
0 & a & 0 \\
a & 0 & b \\
0 & b & 0 
\end{array} \right]
\ ,
\label{eq8}
\end{equation}
$\alpha,~\beta,~a$ and $b$ being real parameters
to be determined from the quark masses.
For simplicity,
these perturbations can be applied separately.
Quark mass matrices (in a flavor basis) are then sums of
the dominant universal FD matrices (6)
plus one kind of the perturbation matrices (8).
One then has to solve the eigenvalue problem
\begin{equation}
\det~|M^{U,D}~-~\lambda|=0,~~{\rm where}~~M^{U,D}=M^{U,D}_0~+~P_i~~
{\rm and}~~\lambda = m_1,~-m_2,~m_3~~
\ ,
\label{eq9}
\end{equation}
and $m_1=m_d~{\rm or}~m_u,~m_2=m_s~{\rm or}~m_c$ and $m_3=m_b$ or $m_t$.
The six parameters of the perturbed matrices $M^{U,D}$
(e.g., $m_t,~{\alpha}^{(u)},~{\beta}^{(u)};~m_b,~a^{(d)},~b^{(d)}$)
are uniquely determined from the experimental input of
the five light (current) quark masses
and the choice of a particular mass for the top quark.
CKM matrix is then constructed as
\begin{equation}
V = U^U_L~
\left[ \begin{array}{ccc}
1 &   &   \\
  & e^{i\sigma} &  \\
  &   &  e^{i\tau}
\end{array} \right]~
U^{D\dagger}_L
\ ,
\label{eq10}
\end{equation}
where phase angles $\sigma$ and $\tau$ are introduced
phenomenologically to generate possible CP violation
in the framework of the three generation standard CKM model.
The CKM matrix is then uniquely determined
by the arbitrary input of the two angles
$\sigma$ and $\tau$ in (10).

To determine these eight perturbation parameters,
a $\chi^2$ analysis was used.
For the first five quarks, the masses obtained by 
Gasser and Leutwyler\cite{5)} can be used.
No constraints on the top quark mass were imposed.
Additional constraints were used -- for four degrees of freedom 
of the CKM matrix coming from two sources.
Information on the quark mixing angles comes
from the measurements of the three absolute values\cite{6)}:
\begin{equation}
|V_{us}| = \sin \theta_{C} = 0.221 \pm 0.002,~~
|V_{cb}| = 0.040 \pm 0.004,~~
\left| V_{ub}/V_{cb} \right| = 0.08 \pm 0.02~~.
\label{eq11}
\end{equation}
Information on the CP violating phase
was taken from the experimental value  of $\varepsilon$ 
parameter of K decay
\begin{equation}
\varepsilon = (2.26 \pm 0.02)~10^{-3} = B_K\cdot f(m_c,m_t,V)
\ ,
\label{eq12}
\end{equation}
where $f$ is a complicated function of the charmed and top quark masses
and of CKM matrix elements, and
$B_K$ is the parameter connecting a free quark estimate to the actual
value of $\Delta S =2$ matrix element describing $K - \bar K$ mixing.
Following Ref.~\cite{8)}, we used the value of $B_K \approx 
2/3~\pm~1/3~~.$

Analysis showed that only the combination of perturbations
$P_U=P_1$ and $P_D=P_2$
resulted in an acceptable value of $\chi^2/d.o.f. \approx 0.6/1$.
The best fit was obtained for
\begin{equation}
m_s = 183~{\rm MeV},~~
m_t = 100~{\rm GeV},~~
\sigma = 0.6^{\circ},~~{\rm and}~~
\tau = 5.7^{\circ}~~,
\label{eq13}
\end{equation}
the other quark masses being close to their central values.
The three other combinations gave much larger values $\chi^2 > 4$.
It appears thus that the prediction 
for the top quark mass from the low energy
FD mass matrices cannot satisfy the TEVATRON\cite{4)} value of
$m_t^{\rm phys.} \approx 175$ GeV.
This model's prediction for
\begin{equation}
J_{CP} = {{\rm Im}}(V_{ub}V_{td}V^*_{ud}V^*_{tb})
\ ,
\label{eq14}
\end{equation}
as a function of $m_t$ can also be obtained --
the approximate value $J_{CP} = (0.3 \pm 0.2)~10^{-4}$ is predicted,
which corresponds to $\sin \delta_{13} \approx (0.56 \pm 0.37)$.
This result is used to predict
\begin{equation}
{\varepsilon'/\varepsilon} = (290)\cdot J_{CP}\cdot H(m_t)
\ ,
\label{eq15}
\end{equation}
where $H(m_t)$ is a decreasing function of the top quark mass\cite{8))}.
The predicted value in the model is
${\varepsilon'/\varepsilon}  = (0.6 \pm 0.5)~10^{-3}$,
with a weak dependence on the top quark mass.
This prediction seems to favor the data from E731\cite{9)}
over the data from NA31\cite{10)}.
\bigskip

To conclude this Section, we described a new set of quark mass matrices
based on a perturbation of a universal (FD) 
Yukawa interaction at {\it low energy\/}.
The model contains eight parameters,
which have been fitted to reproduce the five known quark masses
(except $m_t$),
moduli of three known elements of the CKM matrix,
and the $K$-physics parameter ${\varepsilon}$.
As a result,
the physical top quark mass is predicted to be not much heavier
than $\approx 100$ GeV,
and the direct CP violation parameters are predicted to be
$J_{CP} = (0.3 \pm 0.2)~10^{-4}$
and
${\varepsilon'/\varepsilon}  = (0.6 \pm 0.5)~10^{-3}$.
The analysis will be improved substantially
with a better theoretical knowledge of $B_K$,
a more precise determination of the light quark masses
as well as by taking into account the
more accurate measurement of $|V_{cb}|$ and the ratio
$|V_{ub}/V_{cb}|$.
This {\it low energy\/} model, based on a simple perturbation of  
a universal FD Yukawa interaction at low energies, has been
invalidated by
the discovery of the top quark much heavier than 100 GeV.

\section{Flavor Democracy at High Energies}
Many attempts to unify the gauge interactions of the Standard
Model (SM) have been made in the past --
within the framework of the Grand Unified Theories
(GUT's). These theories give a unification energy 
$E_{\rm GUT} \stackrel{>}{\sim}  10^{16}$ GeV, 
i.e., the energy where the SM gauge couplings
would coincide:
$5 {\alpha}_1/3 =$ $\alpha_2 =$  $\alpha_3$.
Here,  $\alpha_j =  g_j^2 / 4 \pi$ ($j=1,2,3$) are the gauge
couplings of $U(1)_Y,~SU(2)_L$, $SU(3)_C$, respectively. 
For the unification condition 
to be satisfied at a single point $\mu (=E_{\rm GUT})$ exactly,
supersymmetric theories (SUSY) were used,\cite{[1]} replacing the SM 
above the energies $\mu \approx M_{\rm SUSY} \approx 1$ TeV. 
This changed the slopes 
of $\alpha_j=\alpha_j (\mu)$ at $\mu \geq M_{\rm SUSY}$, and 
for certain values of parameters of SUSY the three lines met at a single
point.

There are several deficiencies in such an approach. 
The unification energy is exceedingly large 
($E_{\rm GUT} \stackrel{>}{\sim} 10^{16}$ GeV) since
the proton decay time is large
($\tau_{\rm proton} \geq 5.5 \times 10^{32}$ yr).
This implies a large desert between $M_{\rm SUSY}$ 
and $E_{\rm GUT}$. While eliminating several of the 
previously free parameters of the SM, SUSY introduces several
new parameters and new elementary particles which haven't been
observed.

It is our belief that it is 
more reasonable to attempt first to reduce the number 
of degrees of freedom (d.o.f.'s) in the 
Yukawa sector, since this sector seems to be at least as problematic 
as the gauge boson sector. Any such attempt 
should be required to lead to an overall reduction of the seemingly
independent d.o.f.'s, unlike the GUT--SUSY approach. 
The symmetry responsible for this reduction of the number of parameters 
can be ``flavor democracy'' (FD), valid possibly in certain separate 
sectors of fermions (e.g., up-type sector, down-type sector). 
This symmetry could be realized 
in a flavor gauge theory (FGT)\cite{[2]} -- this
is a theory blind to fermionic flavors at high energies
$E > {\Lambda}_{\rm FGT}$ and leading at
``lower'' energies $E \sim {\Lambda}_{\rm FGT}$ to
flavor--democratic (FD) Yukawa interactions.
Requirement of reduction of as many d.o.f.'s as 
possible would make it natural for FGT's to be without elementary 
Higgs. The scalars of the SM are then tightly
bound states of fermion pairs ${\bar f} f$, 
with ${\bar f}f$ condensation taking place at energies
${\Lambda}$: $E_{\rm ew} \ll {\Lambda} \stackrel{<}{\sim}
{\Lambda}_{\rm FGT}$. 
The idea of FD, and deviations from the exact FD, at {\it low
energies\/} ($E \sim 1-10^2$ GeV) have been investigated 
by several authors\cite{1),[3],11)}. On 
the other hand, in this Section we discuss FD and
deviations from it at {\it higher 
energies\/} $E \gg E_{\rm ew}$, and possible connection with FGT's.
This discussion is motivated and partly based on works of
Ref.~\cite{[2]}.

Let us illustrate first these concepts 
with a simple scheme of an FGT. Assume that
at energies $E \stackrel{>}{\sim} \Lambda_{\rm FGT}$ 
we have no SM scalars, but new gauge bosons $B_\mu$, 
i.e., the symmetry group of the gauge theory is extended to
a group $G_{\rm SM} \times G_{\rm FGT}$. 
Furthermore, we assume that the new gauge bosons obtain a heavy
mass $M_B~( \sim \Lambda_{\rm FGT})$ by an
unspecified mechanism (e.g., dynamically, or via a
mechanism mediated by an elementary Higgs).
At thus high energies,
the SM--part $G_{\rm SM} \equiv$ $SU(3)_c \times SU(2)_L \times U(1)_Y$ 
is without Higgses, and hence with (as yet) massless gauge bosons 
and fermions.
The FGT--part of Lagrangian in the fermionic
sector is written schematically as
\begin{equation}
{\cal{L}}^{\rm FGT}_{g.b.-f} = -g \Psi \gamma^\mu B_\mu \Psi 
~~~({\rm for}~~E \stackrel{>}{\sim} \Lambda_{\rm FGT})~,
\label{eqq2}
\end{equation}
where $\Psi$ is the column of all fermions and $B_\mu=B_\mu^j T_j$. 
$T_j$'s are the generator matrices of the new symmetry group $G_{\rm FGT}$.
Furthermore, we assume that the $T_j$'s corresponding to the electrically
neutral $B_\mu^j$'s do not mix flavors (i.e., no FCNC's at tree level) and 
are proportional to identity matrices in the flavor space
(``flavor blindness''). We will argue in 
the following lines that the FGT Lagrangian (16) can imply creation of
composite Higgs particles through condensation of fermion pairs,
and can subsequently lead at lower energies
to Yukawa couplings with a flavor democracy.

The effective current--current interaction, corresponding to exchanges 
of neutral gauge bosons $B$ at ``low'' cutoff energies $E$ 
($E \sim \Lambda_{\rm FGT} \sim M_B$), is
\begin{equation}
{\cal{L}}^{\rm FGT}_{4f} \approx -{g^2 \over 2 M_B^2} \sum_{i,j} 
(\bar f_i \gamma^\mu f_i)(\bar f_j \gamma_\mu f_j)~~~({\rm for}~~
E \sim \Lambda_{\rm FGT} \sim M_B)~.
\label{eqq3}
\end{equation}
Since we are interested in the possibility of Yukawa interactions of SM 
originating from (17), and since such interactions connect left--handed 
to right--handed fermions, we have to deal only with the left--to--right 
(and right--to--left) part of (17). 
Applying a Fierz transformation\cite{[4]} 
to this part, we obtain four-fermion interactions without
$\gamma_\mu$'s
\begin{equation}
{\cal{L}}^{\rm FGT}_{4f} \approx {2 g^2 \over M_B^2} \sum_{i,j} 
(\bar f_{iL} f_{jR})(\bar f_{jR} f_{iL})~~~({\rm for}~~E \sim
\Lambda_{\rm FGT} \sim M_B)~.
\label{eqq4}
\end{equation}
These interactions can be rewritten in a formally equivalent (Yukawa)
form with auxiliary (i.e., as yet nondynamical) scalar fields. One
possibility is to introduce only one $SU(2)$ 
doublet auxiliary scalar $H$ with 
(as yet arbitrary) bare mass $M_H$, by employing a familiar mathematical 
trick\cite{[5]} 
\begin{eqnarray}
{\cal L}^{(E)}_{\rm Y}& \approx &
- M_H {\sqrt{2} g \over M_B} \sum_{i,j=1}^{3}
{\Bigg \{} \left[ (\bar\psi^q_{iL} \tilde H) u^q_{jR} + (\bar\psi^l_{iL}
\tilde H) u^l_{jR} + \mbox{h.c.} \right]  
\nonumber\\
&&+ \left[ (\bar\psi^q_{iL} H) d^q_{jR} + (\bar\psi^l_{iL} H) d^l_{jR} + 
\mbox{h.c.}
\right] {\Bigg \}} - M_H^2 H^{\dagger} H \  , 
\label{eqq5a}
\end{eqnarray}
where $M_H$ is an unspecified bare mass of the auxiliary
$H$, and we use the notations
\begin{displaymath}
H  =  {H^{+} \choose H^0} \ , \qquad \tilde H  = i \tau_2 H^{\ast} \ ;
\qquad
\psi^q_i = {u^q_i \choose d^q_i} \ , 
\psi^l_i = {u^l_i \choose d^l_i} \ , 
\end{displaymath}
where $u^q_1 = u$, $u_1^l = \nu_e$, $u^q_2=c$, etc.
Another possibility is to  introduce two auxiliary scalar isodoublets 
$H^{(U)},~H^{(D)}$, with (as yet) arbitrary bare masses 
$M_H^{(U)},~ M_H^{(D)}$,  and express (18) in the two-Higgs `Yukawa' form 
\begin{eqnarray}
{\cal{L}}^{(E)}_{\rm Y} 
\approx &- M_H^{(U)} {\sqrt{2} g \over M_B} \sum_{i,j=1}^{3}\left[
(\bar\psi^q_{iL} \tilde H^{(U)}) u^q_{jR} + (\bar\psi^l_{iL} \tilde H^{(U)})
u^l_{jR} + \mbox{h.c.} \right] \nonumber\\ 
 &- M_H^{(D)} {\sqrt{2} g \over M_B} \sum_{i,j=1}^{3} \left[ (\bar\psi^q_{iL}
H^{(D)}) d^q_{jR} + (\bar\psi^l_{iL} H^{(D)}) d^l_{jR} 
+ \mbox{h.c.} \right] \\
&- {M_H^{(U)}}^2 ({H^{(U)}}^\dagger
H^{(U)}) - {M_H^{(D)}}^2 ({H^{(D)}}^\dagger H^{(D)})~~. \nonumber
\label{eqq6}
\end{eqnarray}
The cutoff superscript $E$ ($\sim \Lambda_{\rm FGT}$)
at the ``bare'' parameters and fields in (19) and (20) is
suppressed for simplicity of notation.
Yukawa terms there involve nondynamical 
scalar fields and are formally equivalent to (18). 
Equations of motion show that the (yet)
nondynamical scalars $H$, $H^{(U)}$, $H^{(D)}$ are proportional 
to condensates involving fermions and antifermions -- i.e.,
they are composite.
When further
decreasing the energy cutoff $E$ in the sense of the
renormalization group, 
the composite scalars in (19) and (20) obtain kinetic energy terms
and vacuum expectation values (VEV's) through quantum effects
if the FGT gauge coupling $g$ is strong enough
-- i.e., they become dynamical in an effective SM
(or: two-Higgs-doublet SM) framework and 
they induce dynamically 
electroweak symmetry breaking (DEWSB).
The neutral physical components of these composite Higgs doublets
are scalar condensates\cite{[6]} of fermion pairs 
$H^0 \sim  {\bar f} f$. 
The low energy effective theory is the minimal 
SM (MSM) in the case (19) and the SM with two Higgs doublets --
type II [2HDM(II)] in the case (20). 
Hence, although (19) and (20) are formally equivalent
to four-fermion interactions (18),
they lead to two physically different 
low energy theories\cite{[2]}. The 
condensation scenario with the
smaller vacuum energy density would physically materialize.
We emphasize that the central ingredient distinguishing
the described scheme from most of the other scenarios of
DEWSB is the flavor democracy in the Yukawa sector near the 
transition energies, as expressed in (19) and (20).

We note that (19) implies that the MSM, if it is to be replaced by 
an FGT at high energies, should show up 
a trend of the Yukawa coupling matrix 
(or equivalently: of the mass matrix) in a flavor basis toward a 
complete flavor democracy for {\bf all\/} fermions, 
with a common overall factor, 
as the cutoff energy is increased within 
the effective MSM toward a transition energy
$E_0 (\sim {\Lambda}_{\rm FGT})$
\begin{equation}
M^{(U)}~~{\rm and}~~M^{(D)} \rightarrow {1 \over 3} m_t^0 
\pmatrix{N_{FD}^q & 0 \cr 0 & N_{FD}^l \cr}~~~{\rm as}~~ E \uparrow E_0~~, 
\label{eqq7a}
\end{equation}
where $m_t^0=m_t(\mu=E_0)$ and 
$N_{FD}$ is the $3 \times 3$ flavor--democratic matrix
\begin{equation}
N_{FD}^f = \left[ \begin{array}{ccc}
1 & 1 & 1 \\
1 & 1 & 1 \\
1 & 1 & 1
\end{array} \right]~~,
\label{eqq7b}
\end{equation}
with the superscript $f=q$ for the quark sector and $f=l$ for the leptonic
sector.
On the other hand, if the SM with two Higgses (type II) is to experience such 
a transition, then (20) implies {\bf separate} trends toward FD for the
up--type and down--type fermions
\begin{equation}
M^{(U)}~(M^{(D)}) \rightarrow {1 \over 3}~ m_t^0~(m_b^0)~ 
\left[ \begin{array}{cc}
N_{FD}^q & 0 \\
0 & N_{FD}^l
\end{array} \right]~~~ {\rm as}~~ E \uparrow E_0~~,
\label{eqq8}
\end{equation}
where $m_t^0$ and $m_b^0$ can in general be different.
Note that $N_{FD}$, when written in the diagonal form in the mass
basis, has the form
\begin{equation}
N_{FD}^{\rm mass~basis} = 3 \left[ \begin{array}{ccc}
0 & 0 & 0 \\
0 & 0 & 0 \\
0 & 0 & 1
\end{array} \right]~~.
\label{eqq9}
\end{equation}
Hence, FD (and FGT) implies in the mass basis as $E$ increases to 
$E_0 \sim {\Lambda}_{\rm FGT}$:
\begin{eqnarray}
{m_u \over m_t},~{m_c \over m_t},~{m_{\nu_e} \over m_{\nu_\tau}},~
{m_{\nu_\mu} \over m_{\nu_\tau}} &\rightarrow 0~~, \nonumber\\
{m_d \over m_b},~{m_s \over m_b},~~{m_e \over m_\tau},~~
{m_\mu \over m_\tau} &\rightarrow 0~~, \\
{m_{\nu_\tau} \over m_t},~~{m_\tau \over m_b} &\rightarrow 1~~, \nonumber
\label{eqq10}
\end{eqnarray}
and in the case of the minimal SM {\bf in addition}
\begin{equation}
{m_b \over m_t},~{m_\tau \over m_{\nu_\tau}} \rightarrow 1~~.
\label{eqq11}
\end{equation}
In our previous papers\cite{[2]} we showed, by considering the quark
sector, that the minimal SM does not have the required trend toward FD, but
that SM with two Higgs doublets (type II) does. We also checked that
these conclusions remain true when we include the leptonic sector.
When including also leptons (Ref.~\cite{[2]}, first entry), 
we can neglect for simplicity
masses of the first two families of fermions,
i.e., only $(t,~b)$ and $(\nu_\tau,~\tau)$ are dealt with
(here $\nu_\tau$ is the Dirac tau--neutrino),
and then investigate
evolution of their Yukawa coupling parameters (or: their
masses) with energy.
In the case of the effective 2HDM(II) with
only the third fermion family, 
the FD conditions read as (25) (last line).

The one--loop renormalization group equations (RGE's) 
for the Yukawa coupling parameters
$g_t,~g_b~,g_\nu,~g_\tau$ of the third family fermions in any fixed flavor
basis for various Standard Models with two Higgs doublets are
available, for example, in Ref.\cite{[7]}. 
The running masses (at evolution, or cutoff, energies $E$),
are proportional to these parameters
and to the (running) VEV's of the two Higgs doublets:
\begin{equation}
\left[
\begin{array}{c}
m_t(E) \\
m_{\nu_\tau}(E)
\end{array}
\right] =
\frac{ v_{_U}(E) }{\sqrt{2}} \left[
\begin{array}{c}
g_t(E) \\
g_{\nu_\tau}(E)
\end{array}
\right] \ , \qquad
\left[
\begin{array}{c}
m_b(E) \\
m_{\tau}(E)
\end{array}
\right] =
\frac{ v_{_D}(E) }{\sqrt{2}} \left[
\begin{array}{c}
g_b(E) \\
g_{\tau}(E)
\end{array}
\right] \ , 
\label{eqq12a}
\end{equation}
where
\begin{eqnarray}
\langle H^{(U)(E)} \rangle_0 & = &
{1 \over \sqrt{2}} {0 \choose v_{_U}(E) }~,~~
\langle H^{(D)(E)} \rangle_0 =
{1 \over \sqrt{2}} {0 \choose v_{_D}(E) }~ 
\nonumber\\
{\rm and}~~ v_{_U}^2(E) + v_{_D}^2(E)& = &
v^2(E) \ (\approx 246^2~{\rm GeV}^2 \ \mbox{ for} \ E \sim
E_{\rm ew}) \ .
\label{eqq12b}
\end{eqnarray}
We recall that the transition energy $E_0$,
appearing in FD conditions (25) and (26), 
is the energy above which SM starts being
replaced by an FGT and the composite scalars start
``de-condensing.''
In Ref.\cite{[2]}, we argued that this $E_0$
lies near the pole of the running fermion masses ($E_0
\stackrel{<}{\approx} \Lambda_{\rm pole}$). 
We then simply approximate: $E_0 = \Lambda_{\rm FGT} = \Lambda_{\rm pole}$.
Hence, the high energy boundary conditions (25) are then
\begin{equation}
{g_{\nu_\tau} \over g_t}=1,~~{g_\tau \over g_b}=1~~~ {\rm at}~~E \approx
\Lambda_{\rm pole}~.
\label{eqq13}
\end{equation}
These conditions are taken into account in numerical calculations,
together with the low energy boundary conditions
\begin{eqnarray}
&m_\tau = 1.78~{\rm GeV},~~m_b(\mu=1~{\rm GeV}) = 5.3~{\rm GeV}, \nonumber\\
&m_t(\mu=m_t) \approx 167~{\rm GeV}~, 
\label{eqq14}
\end{eqnarray}
where $m_\tau$ and $m_b$ are based on the available data 
of the measured masses\cite{[8],[9]}. The above value of mass
$m_t(m_t) \approx m_t^{\rm phys.} [1 + 4
{\alpha}_3(m_t)/(3 \pi) ]^{-1}$ $\approx m_t^{\rm phys.}/1.047$ 
is based on the experimental value of $m_t^{\rm phys.}
\approx 175$ GeV measured 
at Tevatron\cite{4)}. For chosen values of
VEV's ratio $v_{_U}/v_{_D}$, we found the masses of Dirac
tau--neutrino $m_{\nu_\tau}$, which satisfy the above boundary conditions 
(29,30), by using numerical integration of RGE's from $\mu=1$ GeV to
$\Lambda_{\rm pole}$.
The calculated Dirac neutrino masses are 
too large to be compatible with results of the available  experimental
predictions\cite{[11]}. Therefore, we invoke the usual 
``see--saw mechanism''\cite{[12]} of
the mixing of the Dirac neutrino masses and the much larger right--handed
Majorana neutrino masses $M_R$, in order
to obtain a small physical  neutrino mass
\begin{equation}
m_\nu^{\rm phys.} \approx {m_\nu^{\rm Dirac} \over 4~ M_R}~~.
\label{eqq15a}
\end{equation}
Majorana mass term breaks the lepton number conservation. Therefore, 
Majorana masses $M_R$ are expected to be of the order of some new
(unification) scale $\Lambda~\gg~E_{\rm ew}$. 
We assume: $M_R \approx \Lambda$.
Within our context, the simplest choice of this new unification scale 
would be the energy $\Lambda_{\rm FGT} = \Lambda_{\rm pole}$ 
where SM is replaced by FGT.
\begin{equation}
m_\nu^{\rm phys.} \approx {m_\nu^{\rm Dirac} \over 4~ \Lambda_{\rm FGT}}~~.
\label{eqq15b}
\end{equation}
The physical tau--neutrino masses $m_{\nu_\tau}^{\rm phys.}$ 
predicted in this
way are very small for the most cases of chosen values 
of $v_{_U}/v_{_D}$ and
$m_t^{\rm phys.}$, i.e., in most cases they are acceptable since being
below the experimentally predicted upper bounds\cite{[11]}. 

The see--saw scenario leading to our predictions of
$m_{\nu_\tau}^{\rm phys.}$ implicitly assumes that:
(a) FGT contains in addition Majorana neutrinos, and its energy range of
validity also provides the scale for the heavy Majorana masses 
[i.e., $M_R \sim \Lambda_{\rm FGT}$]. 
(b) At low (SM) energies, 
Majorana neutrinos remain decoupled from (or
very weakly coupled to) the Dirac neutrinos, which is a very plausible
assumption in view of assumption (a). In general, it could be
assumed $M_R \sim \Lambda_{\rm new-scale} \geq \Lambda_{\rm FGT}$, 
leading thus to even smaller $m_{\nu_\tau}^{\rm phys.}$ 
than those in (32).

When increasing $m_t^{\rm phys.}$ at a fixed
$v_{_U}/v_{_D}$, $m_{\nu_\tau}^{\rm Dirac}$ increases and $\Lambda_{\rm FGT}$
decreases, and hence $m_{\nu_\tau}^{\rm phys.}$ increases. 
This provides us, at a given ratio $v_{_U}/v_{_D}$, with: 
(a) {\it upper\/} bounds on $m_t^{\rm phys.}$ 
for (various) specific upper bounds imposed on 
$m_{\nu_\tau}^{\rm phys.}$ 
(e.g., $\leq 31$  MeV\cite{[11]}, or $\leq 1$ MeV, or
$\leq 17$ KeV\cite{[13]}); (b) {\it lower\/}
bounds on $m_t^{\rm phys.}$ for (various) specific
upper bounds imposed on
$\Lambda_{\rm FGT}$ (e.g.,
$\leq \Lambda_{\rm Planck}$, or 
$\leq 10^{10}$ GeV, or $\leq 10^5$ GeV). Even
with the largest possible upper bounds on $m_{\nu_\tau}^{\rm phys.} \leq 31$ 
MeV and  $\Lambda_{\rm FGT} \leq \Lambda_{\rm Planck}$, we can still get 
rather narrow bands on the values of $m_t^{\rm phys.}$ at any given
$v_{_U}/v_{_D}$. E.g., if $v_{_U}/v_{_D}=1$, then 155 GeV 
$\stackrel{<}{\approx} m_t^{\rm phys.} \stackrel{<}{\approx} 225$ 
GeV. Inversely, if $m_t^{\rm phys.} = 175$ GeV [$m_t(m_t) = 167$
GeV], $m_{\nu_\tau}^{\rm phys.} \leq 31$ MeV and
$\Lambda_{\rm FGT} \leq \Lambda_{\rm Planck}$, then we obtain
rather stringent bounds on the VEV ratio:
$0.64 \stackrel{<}{\approx} v_{_U}/v_{_D} \stackrel{<}{\approx} 1.35$.

To conclude this Section, 
we stress that we can estimate the masses of top 
and tau--neutrino within SM with two Higgs doublets, 
assuming solely that the
complete flavor democracy should set in at energies where SM starts breaking
down. The gauge theories (FGT's) which presumably replace SM at such energies
remain to be further investigated. For related detailed information, see
Ref.\cite{[2]}.

\section{Discussions and Conclusion}
We discussed on the one hand
flavor--democratic (FD) mass
matrices at {\it low energies\/}, and 
on the other hand conditions under which
mass matrices show a trend to flavor--democratic forms 
at {\it high energies\/} (in a flavor basis) -- a behavior 
possibly related to flavor gauge theories (FGT's) at
high energies. 
However, we found that
the model based on our simple perturbation of  
a universal FD Yukawa interaction at {\it low energies\/} has been
invalidated, because of
the discovery of a top quark much heavier than 100 GeV.
On the contrary, at {\it high energies\/}, assuming solely that the
complete flavor democracy should set in at energies where 
an effective perturbative two-Higgs-doublet SM (type II) 
starts breaking down, we can estimate the masses of top 
and tau--neutrino, which are compatible with the present experimental
results. Therefore, the gauge theories (FGT's) which presumably replace 
SM at such energies remain to be further investigated.

In our forthcoming work\cite{chk}, we would like to investigate 
further the simple FD mass matrices ansatz which had been applied
earlier\cite{11)} at low energies and had given
experimentally unacceptable $m_t$. We would like to
apply this ansatz at a high energy scale 
$E \sim \Lambda_{\rm pole}$,
employing RGE evolution within a two-Higgs-doublet SM model
(type II). Furthermore, the compositeness nature of the
scalars in this framework should be further investigated,
particularly in view of the fact that, for cases when VEV
ratio is $v_{_U}/v_{_D} \sim 1$, the usual RGE
compositeness conditions at ${\Lambda}_{\rm pole}$
suggest that only $H^{(U)}$ can be fully composite,
but not $H^{(D)}$ (cf.~Ref.~\cite{rev}).

\section{Acknowledgements}
CSK would like to thank Prof. Y. Koide for his kind invitation 
to the Workshop  of MMQL97.
The work of CSK was supported 
in part by the CTP, Seoul National University, 
in part by Yonsei University Faculty Research Fund of 1997, 
in part by the BSRI Program, Ministry of Education, 
Project No. BSRI-97-2425, and
in part by the KOSEF-DFG large collaboration project, 
Project No. 96-0702-01-01-2. 
The work of GC was
supported in part by the German
Bundesministerium f\"ur Bildung, Wissenschaft,
Forschung und Technologie, Project No. 057DO93P(7).

\end{document}